\documentclass[twoside]{article}
\usepackage{fleqn,espcrc2}

\usepackage{amsmath,amsfonts}

\newcommand{\bra}[1]{\left\langle#1\right|}
\newcommand{\ket}[1]{\left|#1\right\rangle}

\newcommand{\be}{\begin{equation}}
\newcommand{\ee}{\end{equation}}
\newcommand{\bdm}{\begin{displaymath}}
\newcommand{\edm}{\end{displaymath}}
\newcommand{\bea}{\begin{eqnarray}}
\newcommand{\eea}{\end{eqnarray}}
\newcommand{\ba}{\begin{array}}
\newcommand{\ea}{\end{array}}

\newcommand{\1}{{\sf 1 \!\! 1}}

\newcommand{\bn}{\begin{note}}
\newcommand{\en}{\end{note}}

\title{Quantum Link Models with Many Rishon Flavors and with Many
Colors\thanks{Based on a poster by O. B\"ar and a talk by B. Schlittgen.}}
\author{O. B\"ar\address{Center for Theoretical Physics,
        Laboratory for Nuclear Science and Department of 
        Physics,\\
        Massachusetts Institute of Technology (MIT),
        Cambridge, MA 02139, U.S.A.}, %
        R. Brower$^{\mbox{\footnotesize a,}}$\address{Physics Department,
        Boston University, 590 Commonwealth Ave., Boston, MA 02215, U.S.A.},
        B. Schlittgen\address{Center for Theoretical Physics,
        Sloane Physics Laboratory,
        Yale University, P.O. Box 208120,\\
        New Haven, CT 06520, U.S.A.}, %
        U.-J. Wiese$^{\mbox{\footnotesize a,}}$\address{Institute for
        Theoretical Physics,   
        University of Bern,
        Sidlerstra\ss e 5, CH-3012 Bern, Switzerland}
        }

\begin{document}
\begin{abstract}
Quantum link models are a novel formulation of gauge theories in terms
of discrete degrees of freedom. These degrees of freedom are described 
by quantum operators
acting in a finite-dimensional Hilbert space. We show that for certain
representations of 
the operator algebra, the usual Yang-Mills action is recovered in the 
continuum limit. The quantum operators can be expressed as bilinears
of fermionic creation and annihilation operators called rishons.
Using the rishon representation the quantum link Hamiltonian
can  be expressed entirely in terms 
of color-neutral operators. This allows us to
study the large $N_c$ limit of this model. 
In the 't Hooft limit we find an area 
law for the Wilson loop and a mass gap. Furthermore, the strong coupling
expansion is a topological expansion in which graphs with handles and
boundaries are suppressed. 
\vspace{1pc}
\end{abstract}

\maketitle

\section{Introduction}
The quantum link formulation of lattice gauge theories is an 
approach which is very different from
the standard framework based on a Euclidean path integral
\cite{Wiese:1999ab}. Since quantum links are discrete variables,
i.e.\ operators that act in a finite dimensional
Hilbert space, it is hoped that the quantum link formulation will allow
numerical simulations using cluster algorithms. 
Meron-cluster algorithms
\cite{Chandrasekharan:1999cm,Cox:1999nt}, for instance,
proved to be superior to standard methods for a wide range of
models, such as the 2-d O(3) model with $\theta=\pi$ and with a nonzero
chemical potential $\mu$, the Potts model of QCD, and various
fermionic systems with a sign problem, like the attractive Hubbard
model. 

Furthermore quantum link models have several features
that go beyond the standard formulation. For example, a quantum link
can be written as a bilinear of fermionic creation and annihilation
operators \cite{Brower:1997ha},
\begin{equation}
\hat{U}_{x,\mu}^{ij}\,=\,
\hat{c}_{x,\mu}^{i\alpha}\,\hat{c}_{x+\hat{\mu},-\mu}^{j\alpha\dagger}\,.
\end{equation}
These fermions are  called rishons. This splitting of the link
variable has no analogue in Wilson's formulation of lattice QCD.  
In the following, we make use of the rishon representation to investigate the
large $N_c$ limit of the quantum link model. Using coherent
state techniques we derive a new effective theory for 
large $N_c$ gauge theory.   
 
\section{The quantum link model} 

The quantum link model is described by a Hamilton operator $\hat{H}$, which
evolves the system in an additional fifth dimension
\cite{Chandrasekharan:1997ih,Brower:1997ha}. 
The partition function and correlation functions are given by
operator traces, e.g.\ 
$\label{QL_Z}
{\cal Z} \,=\, \mbox{Tr }e^{-\beta \hat{H}}$,
where $\beta$ is the extent of the additional dimension.
This extra dimension should eventually disappear via dimensional reduction,
thereby giving rise to the 4-d target theory.
In the next section, we will discuss under what circumstances this
will happen.

The Hamilton operator is obtained from Wilson's plaquette action by 
replacing the elements of $SU(N_c)$ link matrices --- ordinary
complex numbers --- by operators acting in a finite-dimensional
Hilbert space,
\begin{equation}
\label{eq:Hamiltonian}
\hat{H}\,=\,J\sum_{x,\,\mu\neq\nu} \mbox{tr}\,\left(\hat{U}_{x,\mu}
\hat{U}_{x+\hat{\mu},\nu}\hat{U}^{\dagger}_{x+\hat{\nu},\mu}
\hat{U}^{\dagger}_{x,\nu}\right).
\end{equation}
Here, the trace sums only over the color indices. 
The commutation relations of the quantum link operators are determined
by the requirement that the Hamilton operator (\ref{eq:Hamiltonian}) be
invariant under $SU(N_c)$ gauge transformations. 

Gauge invariance is expressed as $[\hat{H},\hat{G}_x^a]=0$.
The generators of gauge transformations satisfy $[\hat{G}_x^a,\hat{G}_y^b]=
2if^{abc}\hat{G}_x^c\delta_{x,y}$, where $f^{abc}$ are the
structure constants of $SU(N_c)$.
Writing  $\hat{G}_x^a
\,=\,\sum_{\mu}\big(\hat{L}_{x,\mu}^a + 
\hat{R}_{x-\hat{\mu},\mu}^a \big)$,
we find the following commutation relations between
$\hat{R},\hat{L},\hat{U}$ and $\hat{U}^{\dagger}$ ($x$ and $\mu$ 
indices suppressed):
\begin{align}
[\hat{R}^a,\hat{R}^b]&=\,2if^{abc}\hat{R}^c,
&[\hat{L}^a,\hat{L}^b]&=\,2if^{abc}\hat{L}^c, \nonumber\\
[\hat{R}^a,\hat{U}]&=\hat{U}\lambda^a,
&[\hat{L}^a,\hat{U}]&=-\lambda^a \hat{U}, \nonumber\\
[\hat{R}^a,\hat{U}^{\dagger}]&=\hat{U}^{\dagger}\lambda^a,
&[\hat{L}^a,\hat{U}^{\dagger}]&=-\lambda^a \hat{U}^{\dagger}.
\end{align}
Here, the $\lambda^a$ are $SU(N_c)$ generators in the defining
representation.
The commutation relations between the $\hat{U}$'s and $\hat{U}^{\dagger}$'s are
as yet undetermined. The smallest algebra under which the
above relations close is $SU(2N_c)$, provided we
{\em postulate} the 
proper commutation relations between the $\hat{U}$'s and
$\hat{U}^{\dagger}$'s. Hence, 
an $SU(2N_c)$ algebra is associated with each link. 
Actually, this construction leads to a $U(N_c)$ gauge theory, and
in order to break the unwanted $U(1)$ symmetry and obtain an
$SU(N_c)$ gauge theory, one can add a
determinant term to $\hat{H}$, 
\begin{equation}
\label{eq:det}
\hat{H}'=J'\sum_{x,\mu}\left[ \det\hat{U}_{x,\mu}+
\det\hat{U}_{x,\mu}^{\dag}\right],
\end{equation}
thus breaking the unwanted $U(1)$ \cite{Brower:1997ha}.

The $SU(2N_c)$ generators may be expressed as products of fermionic
creation and annihilation operators, which obey canonical 
anti-commutation relations:
\begin{align}
\label{eq:generators}
\hat{U}^{ij}_{x,\mu}&=
\hat{c}_{x,\mu}^{i\alpha}\hat{c}_{x+\hat{\mu},-\mu}^{j\alpha\dagger}, 
&\hat{U}^{ij\dagger}&=\hat{c}_{x+\hat{\mu},-\mu}^{j\alpha}
\hat{c}_{x,\mu}^{i\alpha\dagger},\\
\hat{L}^a_{x,\mu} &=
\hat{c}_{x,\mu}^{i\alpha\dagger}\lambda^a_{ij}\hat{c}_{x,\mu}^{j\alpha},
&\hat{R}^a_{x,\mu}  &=
\hat{c}_{x+\hat{\mu},-\mu}^{i\alpha\dagger}\lambda^a_{ij}
\hat{c}_{x+\hat{\mu},-\mu}^{j\alpha}. \nonumber
\end{align}
These fermions are called rishons, and in addition to the color index
$i$ they carry a rishon flavor index $\alpha$, which runs from $1$ to $M$.
In fact, their introduction
corresponds to the choice of a particular representation for the 
$SU(2N_c)$ algebra, represented by a rectangular Young tableau with 
$N_c$ rows and $M$ columns,
%
%
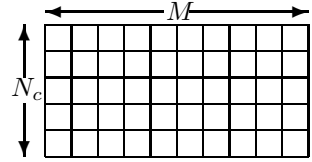
\begin{center}
\setlength{\unitlength}{0.7mm}
\begin{picture}(55,30)
\multiput(5,0)(5,0){11}{\line(0,1){25}}
\multiput(5,0)(0,5){6}{\line(1,0){50}}

\put(28,26){$M$}
\put(-1.5,11.5){$N_c$}

\thicklines
\put(1,10){\vector(0,-1){10}}
\put(1,15){\vector(0,1){10}}

\put(27.5,27.5){\vector(-1,0){22.5}}
\put(32.5,27.5){\vector(1,0){22.5}}
\end{picture}\end{center}
%
%
%
%
%
We always choose the representation with vertical length $N_c$. This
so-called half-filling constraint is equivalent to
fixing the total number of rishons on a link to be $N_cM$,
\begin{equation}
\sum_i\left(\hat{c}_{x,\mu}^{i\alpha\dagger}\hat{c}_{x,\mu}^{i\beta}
+\hat{c}_{x+\hat{\mu},-\mu}^{i\alpha\dagger}
\hat{c}_{x+\hat{\mu},-\mu}^{i\beta} \right)=\delta^{\alpha\beta}N_c.
\end{equation}
For these representations the determinant term in (\ref{eq:det})
is non-trivial and hence the unwanted $U(1)$ symmetry is broken.

\section{The low-energy effective theory}

The physics of the quantum link model is completely determined, once
the representation for the $SU(2N_c)$ quantum link algebra has been
chosen. Having picked a particular type of representation in the
previous section, we would like to investigate the dynamics of the
theory. In particular, we are interested in determining whether the
theory undergoes dimensional reduction and what kind of effective
theory is described in the long-distance limit. To this end, we start 
by setting up 
a coherent state path integral. Letting $M$ get large, this
will allow us to do a semi-classical expansion around the classical
ground state, whence we recover Yang-Mills theory as a low-energy
effective action of the theory \cite{Schlittgen:2001xg}. 
Generalized coherent states for $SU(2N_c)$
are generated by acting with all group elements on the highest weight
state of the chosen representation \cite{Perelomov:1986,Read:1989jy}.
In our case, the highest weight state is given by
\begin{equation}
\ket{\psi_0}=\mathcal{C}\left[\epsilon^{ab\cdots}c^{a\alpha\dag}
c^{b\alpha\dag}\cdots \right]^{M}\ket{0},
\end{equation}
where there are $N_c$ creation operators inside the square bracket. The indices
$a,b,\ldots$ run from 1 to $N_c$, and $\alpha$ runs from
1 to $M$. Collectively denoting the $SU(2N_c)$ generators of 
eq.(\ref{eq:generators}) by $\hat{S}^{ij}$,
coherent states are given by
\begin{equation}
\ket{q}=\exp\left( -q^{ij}\hat{S}^{ji}+q^{ij*}\hat{S}^{ij}\right)
\ket{\psi_0},
\end{equation}
for $1\leq j\leq N_c$ and $N_c+1\leq i\leq 2N_c$. 
These states have a number of nice properties. Most importantly,
the identity operator can be resolved as
\begin{equation}
\1 = \int dq \ket{q}\bra{q}.
\end{equation}
Furthermore, we have the following expression for diagonal matrix elements
of $SU(2N_c)$ generators between coherent states,
\begin{equation}
\bra{q}\hat{S}^{ij}\ket{q}= (M/2) Q^{ij},
\end{equation}
where the matrix $Q$ parameterizes the coset space $SU(2N_c)/[SU(N_c)\times
SU(N_c)\times U(1)]$,
\begin{equation}
Q = \exp\left(
\begin{array}{@{}cc@{}}
0&q^{\dag}\\
-q&0
\end{array}\right)
\left(
\begin{array}{@{}cc@{}}
1_N&0\\
0&-1_N
\end{array}\right)
\exp\left(
\begin{array}{@{}cc@{}}
0&-q^{\dag}\\
q&0
\end{array}\right).
\end{equation}
Note that this coset space arises, because the stability subgroup
of $\ket{\psi_0}$ is $SU(N_c)\times SU(N_c)\times U(1)$.
Using these properties, it is straightforward to set up a coherent
state path integral in the standard way, which is an integral over
$GL(N,\mathbb{C})$ matrices $q$. 

In order to derive the low-energy effective theory, we do a semi-classical
expansion around the classical ground state. The validity of this
expansion depends on the existence of an intermediate momentum
cut-off scale, which is much larger than the inverse correlation
length and much smaller than the inverse lattice spacing, 
$\xi^{-1}\ll \Lambda\ll a^{-1}$. Then, 
dominant contributions to the path integral are slowly varying
on this intermediate scale, 
and higher order terms can be ignored. Below, we will
see that such a cut-off scale exists for large enough values of $M$,
i.e.\ large enough representations of the quantum link algebra.
To expand around the minimum of the action,
we decompose the matrices $q=bu$ into a Hermitian part $b$ and
a unitary part $u$, and write
\begin{equation}
u_{\mu}=\exp(-ia^2 (\theta_{\mu}/N)\1-iaA_{\mu}),
\end{equation}
and $b_{\mu} = (\pi/4)\1 + a^2 E_{\mu}$, and expand in powers of the lattice
spacing $a$, keeping only terms up to order $a^4$. We then 
integrate out the quadratic fluctuations $E^2$, to obtain
an effective five-dimensional action,
\begin{equation}
\label{eq:5-d_ac}
S[A] = \!\int_0^{\beta}\!\!dx_5\int \!d^4x \frac{1}{4e^2}
\left(\mbox{tr}\,F_{\mu\nu}^2 +
\frac{1}{c^2}\mbox{tr}\,(\partial_5 A_{\mu})^2\right)
\end{equation}
Here, $e=8/(M^4 J)$ is the 5-d gauge coupling, and
$c=(Ma/2)\sqrt{\gamma J}$ is the ``velocity of light''.
The constant $\gamma$ is given by $\gamma = 3JM^4 +4J'(M/2)^{N_c}$.
Notice that there is no $A_5$ field in the action. This is due
to the fact, that we did not impose Gauss' law, as we have no need
for 5-dimensional gauge invariance. The 4-dimensional gauge
invariance is intact, since $A_4\neq 0$.
Taking the extent $\beta$ of the fifth dimension to be infinite, 
we nevertheless have a 5-dimensional gauge theory in $A_5=0$ gauge.
Note that
a five-dimensional gauge theory can exist in a non-Abelian Coulomb phase.
This is generically the case for lattice gauge theories such as ours
\cite{Creutz:1979dw,Beard:1997ic}, and
this phase is characterized by massless gluons $A_{\mu}$
and hence an infinite correlation length $\xi$. When $\beta$ is made
finite, the confinement hypothesis requires 
that gluons pick up a non-perturbatively generated mass.
However, $\xi$ is still exponentially larger than $\beta$,
\begin{equation}
\label{eq:corr_len}
\xi\propto \exp[24\pi^2\beta/(11N_c e^2)],
\end{equation}
and the theory is dimensionally reduced to 4 dimensions,
\begin{equation}
S[A] = \int d^4x \,\frac{\beta}{4e^2}\,
\left(\mbox{tr}\,F_{\mu\nu}^2 \right)\,+\,\ldots
\end{equation}
This is the standard gauge field action with the 4-d gauge coupling $g$
given by
$1/g^2\,=\,\beta/e^2$.
Thus, the continuum limit $g\to 0$ is approached by sending the
extent $\beta$ to infinity, and
we see that the low-energy dynamics of the quantum link model
describes ordinary 4-d Yang-Mills theory.

Notice that the dependence of $e^2$ on $M$ leads to a large
correlation length (\ref{eq:corr_len}) for large $M$.
At fixed $\beta$, this guarantees the existence of an intermediate momentum
cut-off scale mentioned above, if we choose $M$ large enough, and
hence, the semi-classical expansion is valid in this case.

%
\section{The quantum $\hat{\Phi}$--model and large $N_c$}
%

Consider the Hamilton operator of the $U(N_c)$ quantum link
model in (\ref{eq:Hamiltonian}), without the determinant term, i.e.\
$J'=0$. Writing the link operators in terms of rishons, 
the product of four quantum links becomes 
a product of eight rishon operators. 
We can thus group together rishon operators at each lattice point 
to form $\hat{\Phi}$-operators:
\begin{equation}
\hat{\Phi}^{\alpha\beta}_{x,\mu,\nu}\,=\,\sum_i
\hat{c}_{x,\mu}^{i\alpha}\hat{c}_{x,\nu}^{i\beta\dagger}\,.
\end{equation}
These operators transform as color singlets since color
indices are summed over. In terms of
these operators the Hamilton operator reads \newline

\noindent$\hat{H}= -J\times$
\begin{equation}
\sum_{x,\mu\neq\nu}
\mbox{tr}\big(\hat{\Phi}_{x,\nu,\mu}
\hat{\Phi}_{x+\hat{\mu},-\mu,\nu}
\hat{\Phi}_{x+\hat{\mu}+\hat{\nu},-\nu,-\mu}
\hat{\Phi}_{x+\hat{\nu},\mu,-\nu}\big)\,.
\end{equation}
Here the trace sums over rishon flavor indices. The sign appears
because we have to commute the last  -- fermionic -- rishon operator
an odd number of times.

The Hamiltonian in terms of the $\hat{\Phi}$'s has a
local $U(M)$ flavor symmetry. The color symmetry, on the other hand, 
is completely hidden in the
definition of the $\hat{\Phi}$--operators.

As already mentioned, a
$U(2N_c)$ algebra is associated with each link of the lattice when 
working with the quantum link operators $\hat{U},\hat{U}^{\dagger}$.
Using the anti-commutation relations between the rishon operators one can
work out the commutation relations between the
$\hat{\Phi}$-operators. One
finds that they generate a $U(2dM)$ algebra at each lattice point, where
$d$ is the number of physical space-time dimensions.

Moreover, fixing the number of rishons at each lattice site to be
$dMN_c$ (recall
half-filling) is
equivalent to choosing a representation for $U(2dM)$ with the
rectangular Young-tableau
%
%
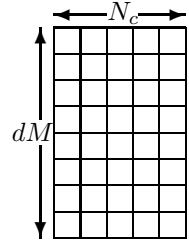
\begin{center}
\setlength{\unitlength}{0.7mm}
\begin{picture}(30,45)
\multiput(5,0)(5,0){6}{\line(0,1){40}}
\multiput(5,0)(0,5){9}{\line(1,0){25}}

\put(15.2,41.5){$N_c$}
\put(-3,18.5){$dM$}

\thicklines
\put(2.5,17.5){\vector(0,-1){17.5}}
\put(2.5,22.5){\vector(0,1){17.5}}

\put(15,42.5){\vector(-1,0){10}}
\put(21,42.5){\vector(1,0){9}}
\end{picture}\end{center}
What is the advantage of working with the $\hat{\Phi}$-operators? 
By using the $\hat{\Phi}$-operators
the r\^{o}les of $M$ and $N_c$ are reversed: 
the number of colors $N_c$ determines the size of the
representation of $U(2dM)$. In the last section, we 
derived a 4-d low-energy effective theory of the quantum link model for
large representations. We recover 4-d Yang-Mills theory for large $M$. 
This suggests the use of similar methods to
derive an effective theory for the $\hat{\Phi}$-model and obtain
the large $N_c$ limit of pure gauge theory! 

It turns out that one can indeed take
the limit $N_c\rightarrow\infty$ in the $\hat{\Phi}$--model for any
value of $M$. But before getting too excited about this, one should bear
in mind that in the 't Hooft limit of gauge theories 
$g^2N_c$ is kept fixed as $N_c\rightarrow\infty$. 

To determine what this implies for $M$,
recall the 5-d effective action describing the
low-energy dynamics of the quantum link model (\ref{eq:5-d_ac}).
As already pointed out, the parameters $e$ and $c$ of the five
dimensional theory are functions of $N_c$ and $M$ and determine the
parameters of the dimensionally reduced 4-d theory.

The 't Hooft limit of the 5-d theory is given by

$e^2 N_c=\mbox{const.}$ and
$c=\mbox{const.}$ as $N_c\rightarrow \infty$.
Unfortunately, this implies that $M\propto N_c$,
and so the 't Hooft limit requires $M\rightarrow\infty$ as well. 
%
\section{The classical $\phi$-model}
%
The classical $\phi$-model is the 4-d effective theory
of the quantum  $\hat{\Phi}$-model in the limit $N_c\rightarrow \infty$,
and after dimensional reduction.  
It's derivation is analogous to the derivation of the effective
theory of the original quantum link model, using
coherent state techniques. We skip the details concerning the coherent
states and present directly the resulting 4-d lattice theory. 

The 4-d $\phi$-model assigns a field variable
\begin{equation}
\phi_{\mu,\nu}(x)\in GL(M,\mathbb{C})\,,\qquad\mu,\nu\,=\,\pm 1,\ldots\,
\pm 4
\end{equation}
to each lattice point. The Greek indices can take positive and
negative values. The negative index $-\mu$
refers to the link 'pointing' in the opposite $\mu$ direction.

The field variables $\phi_{\mu,\nu}(x)$ at a lattice point are
not independent. They are constrained to be elements of the 
coset space $U(2dM)/[U(dM)\times U(dM)]$,
\begin{equation}\label{representation_phi}
\phi \,=\,U\,J\,U^{\dagger}\,,
\end{equation}
where $U$ is an element of $U(2dM)$ and $J$ being the diagonal matrix
$\mbox{diag}(1,\ldots, 1,-1,\ldots,-1)$ 
with $dM$ 1's and -1's, respectively. 

The action contains two parts. The first term reads 
\begin{equation}\label{Phi_action_1}
S_1[\phi] \, = \, \beta\sum_{x,\,\mu\neq\nu} \mbox{tr}\, 
P_{\mu\nu}(x)\,+\,\mbox{h.c.}
\end{equation}
with $P_{\mu\nu}$ defined as the product of four
$\phi$'s at the corners of the basic plaquettes, explicitly:
\begin{eqnarray}\label{def_plaquette}
P_{\mu,\nu}(x) &=&
\phi_{\mu,\nu}(x)\phi_{-\nu,\mu}(x+a\hat{\nu})\,\times\\
 & &\phi_{-\mu,-\nu}
(x+a\hat{\nu}+a\hat{\mu})\phi_{\nu,-\mu}(x+a\hat{\mu}). \nonumber
\end{eqnarray}
This term directly
stems from the plaquette term in the Hamiltonian of the quantum link
model. 

The half-filling constraint for the $U(2dM)$ representations of the
quantum $\hat{\Phi}$ model results in a constraint for the    
$\phi$-fields at neighboring lattice
points,
\begin{equation}\label{half_filling_constraint}
\sum_{d\geq\mu\geq 1}\Big(\phi_{\mu,\mu}(x) +
\phi_{-\mu,-\mu}(x+a\hat{\mu})\Big)\,=\,0\,.
\end{equation}
This constraint is encouraged by adding a second part to the gauge
action, 
\begin{equation}\label{Phi_action_2}
S_2[\Phi] \, = \, 
\lambda\sum_{x,\mu}\mbox{tr}\left[\phi_{\mu,\mu}(x) +
\phi_{-\mu,-\mu}(x+a\hat{\mu})\right]^2\,.
\end{equation}
This $\lambda$-term 
imposes a soft version of the half-filling constraint.

A crucial
feature of the total action $S_1+S_2$ is its invariance under local
$U(M)$ rotations. 
It is obvious from (\ref{Phi_action_1}) and (\ref{Phi_action_2}) that
the action is invariant under the transformation
\begin{equation}
\phi_{\mu,\nu}(x) \,\longrightarrow \,
g_{\mu}(x)\phi_{\mu,\nu}(x)g_{\nu}(x)^{-1}\,, 
\end{equation}
if we define $g_{-\nu}(x)\,=\, g_{\nu}(x-a\hat{\nu})$ 
for the non-positive values $-1,\ldots,-d$ of $\nu$.

To complete our definition of the $\phi$-model we need to specify an
integration measure for the functional integral. In view
of (\ref{representation_phi}) it is natural to define 
\begin{equation}\label{def_measure}
\mbox{D}[\phi]\, = \,\prod_{x}\mbox{d}\phi(x)\,,
\end{equation}
where $\mbox{d}\phi(x)$ is the Haar measure of the group $U(2d M)$. 
Obviously this measure is invariant under gauge transformations.

With the measure at hand we define the partition function
\begin{equation}\label{def_partition_function}
Z\,=\,\int\mbox{D}[\phi]\,e^{-S[\phi]}
\end{equation}
and correlation functions 
\begin{equation}\label{def_correlation_functions}
\langle{\cal O}\rangle\,=\,\frac{1}{Z}\int\mbox{D}[\phi]\,{\cal
O}e^{-S[\phi]}
\end{equation}
of any product ${\cal O}$ of the $\phi$-fields. 

As discussed before, the $\phi$-model we have defined
here should be equivalent to large $N_c$ pure gauge theory in the limit
$M\rightarrow\infty$. Sending $M$ to infinity, however, we have to keep 
\begin{equation}
\beta'\,=\,\frac{\beta}{M}\, ,\qquad \lambda'\,=\,\frac{\lambda}{M}\, ,
\end{equation}
finite, similar to the 't Hooft limit. 
\section{Strong coupling results}
As a first step in the investigation of the $\phi$-model, 
we looked at its strong coupling expansion.
In this expansion the Boltzmann
weight $e^{-S[\phi]}$ in the functional integral is expanded in powers
of $\beta$ and $\lambda$. This leads to a power series expansion of
the observable for any {\em finite} $M$. In the end we haven taken the
infinite $M$ limit. We find the following
results:
\vspace{0.2cm}  

1. At strong coupling the $\phi$-model has a mass gap.
In the limit
$M\rightarrow\infty$ the lowest glue ball mass at leading order is given by
\begin{displaymath}
m\,=\, -2\ln 2\frac{\beta'}{8^3}\,.
\end{displaymath}
\indent 2. An area law is found for the Wilson loop 
and in the limit
$M\rightarrow\infty$ the string tension at leading order is given by
\begin{displaymath}
\sigma\,=\, -\ln \frac{\beta'}{8^3}\,.
\end{displaymath}
This result is essentially a consequence of the $U(M)$ rishon flavor symmetry.
To obtain a non-zero expectation value, the expansion of the Boltzmann factor
needs to generate at least the plaquettes inside the Wilson loop. 
This is analogous to
the strong coupling expansion in conventional pure lattice gauge
theory.

3. The strong coupling expansion is a topological expansion. 
Strong
coupling graphs with $H$ handles and $B$ boundaries (or 
quark loops) are suppressed by inverse
powers of $M$, 
\begin{equation}
\label{top_expansion}
(1/M)^{2H + B - 2}\,. 
\end{equation}
To leading order only planar  diagrams contribute.\\[0.2ex]

From these results we conclude that the $\phi$-model shares some of the
qualitative features of large $N_c$ gauge theory, 
known from the conventional
approach \cite{'tHooft:1974jz,'tHooft:1974hx,Witten:1979kh}. 
Note, however, that it seems almost impossible to
come up with the $\phi$-model as a reformulation of large $N_c$ gauge theory 
without the detour via the quantum link formulation. 
%
\section{Conclusions and Outlook}
%

In conclusion, we showed analytically, that ordinary 4-d Yang-Mills theory
is the low-energy effective theory of the quantum link model, provided the 
representation of the quantum link algebra is sufficiently large.
Hence the quantum link approach is a valid formulation of gauge field theory.

The large $N_c$ limit of quantum link models can be investigated
in terms of color singlet fields.
This so-called $\phi$-model exhibits important
qualitative features of large $N_c$ pure gauge theory like a mass gap and an
area law for the Wilson loop.

There are several directions for future
research. It would be interesting to study the $\phi$-model at
weak coupling and compute the $\beta$-function to see if 
this model is asymptotically free, as expected. 
Furthermore, numerical simulations of the $\phi$-model for various 
values of $M$ may lead to further insight into the model. It is quite
possible, that fairly small values of $M$ are
already suffient for studying the large $N_c$ limit. 
In order to study full QCD, fermions should be included.
By pairing rishon and quark operators, the theory can be completely
bosonized \cite{Brower:1997ha} which would be a significant advantage in
numerical simulations.

This work is supported in part by funds provided by the U.S Department
of Energy (D.O.E.) under cooperative agreements DE-FC02-94ER40818
and DE-FC02-91ER40676.



\end{document}